%% file: main.tex
\documentclass[reprint, superscriptaddress, prl, floatfix]{revtex4-2}
\usepackage{amsmath,amssymb}
\usepackage{graphicx}
\usepackage{bm}
\usepackage{times}
\usepackage{xcolor}
\usepackage{lineno}
\usepackage{textcomp, gensymb}
\usepackage{hyperref}
\usepackage{tabularx}
\usepackage{adjustbox}
\usepackage{xspace}
\usepackage{color,soul}
\usepackage{rotating}

\newcommand{\nuebar}{\ensuremath{\overline{\nu}_e}\xspace}
\newcommand{\uFive}{$^{235}$U}

\newcommand{\NIST}{National Institute of Standards and Technology (NIST) \renewcommand{\NIST}{NIST}}
\newcommand{\NBSR}{National Bureau of Standards Reactor (NBSR) \renewcommand{\NBSR}{NBSR}}
\newcommand{\INL}{Idaho National Laboratory (INL) \renewcommand{\INL}{INL}}
\newcommand{\ATR}{Advanced Test Reactor (ATR) \renewcommand{\ATR}{ATR}}
\newcommand{\ORNL}{Oak Ridge National Laboratory (ORNL) \renewcommand{\ORNL}{ORNL}}
\newcommand{\HFIR}{High Flux Isotope Reactor (HFIR) \renewcommand{\HFIR}{HFIR}}
\newcommand{\LLNL}{Lawrence Livermore National Laboratory (LLNL) \renewcommand{\LLNL}{LLNL}}

\usepackage{verbatim}

\begin{document}


\title{Final Search for Short-Baseline Neutrino Oscillations with the PROSPECT-I Detector at HFIR}

\input{AuthorList2022.tex}

\collaboration{The PROSPECT Collaboration}
\homepage{http://prospect.yale.edu \\ email: prospect.collaboration@gmail.com}

\begin{abstract}
The PROSPECT experiment is designed to perform precise searches for antineutrino disappearance  at short distances (7 -- 9~m) from compact nuclear reactor cores.
This Letter reports results from a new neutrino oscillation analysis performed using the complete data sample from the PROSPECT-I detector operated at the High Flux Isotope Reactor in 2018.  
The analysis uses a multi-period selection of inverse beta decay neutrino interactions with reduced backgrounds and enhanced statistical power to set limits on electron neutrino disappearance caused by mixing with sterile neutrinos with 0.2 -- 20 eV$^2$ mass splittings.  
Inverse beta decay positron energy spectra from six different reactor-detector distance ranges are found to be statistically consistent with one another, as would be expected in the absence of sterile neutrino oscillations.  
The data excludes at 95\% confidence level the existence of sterile neutrinos in regions above 3~eV$^2$ previously unexplored by terrestrial experiments, including all space below 10~eV$^2$ suggested by the recently strengthened Gallium Anomaly.  
The best-fit point of the Neutrino-4 reactor experiment's claimed observation of short-baseline oscillation is ruled out at more than five standard deviations.  
\end{abstract}

\maketitle



\input{Introduction.tex}
\input{ExperimentalDescription.tex}

\input{AnalysisMethods.tex}

\input{Results.tex}


This material is based upon work supported by the following sources: US Department of Energy (DOE) Office of Science, Office of High Energy Physics under Award No. DE-SC0016357 and DE-SC0017660 to Yale University, under Award No. DE-SC0017815 to Drexel University, under Award No. DE-SC0008347 to Illinois Institute of Technology, under Award No. DE-SC0010504 to University of Hawaii, under Contract No. DE-SC0012704 to Brookhaven National Laboratory, and under Work Proposal Number  SCW1504 to Lawrence Livermore National Laboratory. This work was performed under the auspices of the U.S. Department of Energy by Lawrence Livermore National Laboratory under Contract DE-AC52-07NA27344 and by Oak Ridge National Laboratory under Contract DE-AC05-00OR22725. Additional funding for the experiment was provided by the Heising-Simons Foundation under Award No. \#2016-117 to Yale University.
 
 
We further acknowledge support from Yale University, the Illinois Institute of Technology, Temple University, University of Hawaii, Brookhaven National Laboratory, the Lawrence Livermore National Laboratory LDRD program, the National Institute of Standards and Technology, and Oak Ridge National Laboratory. We gratefully acknowledge the support and hospitality of the High Flux Isotope Reactor and Oak Ridge National Laboratory, managed by UT-Battelle for the U.S. Department of Energy.

\bibliography{references}


\end{document}

%% file: AuthorList2022.tex
\address{Department of Physics, Boston University, Boston, MA, USA} \vspace{-0.4\baselineskip}
\address{Brookhaven National Laboratory, Upton, NY, USA} \vspace{-0.4\baselineskip}
\address{Department of Physics, Drexel University, Philadelphia, PA, USA} \vspace{-0.4\baselineskip}
\address{George W.\,Woodruff School of Mechanical Engineering, Georgia Institute of Technology, Atlanta, GA, USA} \vspace{-0.4\baselineskip}
\address{Department of Physics and Astronomy, University of Hawaii, Honolulu, HI, USA} \vspace{-0.4\baselineskip}
\address{Department of Physics, Illinois Institute of Technology, Chicago, IL, US} \vspace{-0.4\baselineskip}
\address{Nuclear and Chemical Sciences Division, Lawrence Livermore National Laboratory, Livermore, CA, USA} \vspace{-0.4\baselineskip}
\address{Department of Physics, Le Moyne College, Syracuse, NY, USA} \vspace{-0.4\baselineskip}
\address{National Institute of Standards and Technology, Gaithersburg, MD, USA} \vspace{-0.4\baselineskip}
\address{High Flux Isotope Reactor, Oak Ridge National Laboratory, Oak Ridge, TN, USA} \vspace{-0.4\baselineskip}
\address{Physics Division, Oak Ridge National Laboratory, Oak Ridge, TN, USA} \vspace{-0.4\baselineskip}
\address{Department of Physics, Susquehanna University, Selinsgrove, PA, USA} \vspace{-0.4\baselineskip}
\address{Department of Physics and Astronomy, University of Tennessee, Knoxville, TN, USA} \vspace{-0.4\baselineskip}
\address{Department of Physics, United States Naval Academy, Annapolis, MD, USA} \vspace{-0.4\baselineskip}
\address{Department of Physics, University of Wisconsin, Madison, WI, USA} \vspace{-0.4\baselineskip}
\address{Wright Laboratory, Department of Physics, Yale University, New Haven, CT, USA} \vspace{-0.4\baselineskip}

\author{M.~Andriamirado$^{6}$,
A.~B.~Balantekin$^{15}$,
C.~D.~Bass$^{8}$,
O. Benevides Rodrigues$^{6}$,
E.~P.~Bernard$^{7}$,
N.~S.~Bowden$^{7}$,
C.~D.~Bryan$^{10}$,
R.~Carr$^{14}$, 
T.~Classen$^{7}$,
A.~J.~Conant$^{10}$,
G.~Deichert$^{10}$,
M.~J.~Dolinski$^{3}$,
A.~Erickson$^{4}$,
A.~Galindo-Uribarri$^{11,13}$,
S.~Gokhale$^{2}$,
C.~Grant$^{1}$, 
S.~Hans$^{2}$,
A.~B.~Hansell$^{12}$,
K.~M.~Heeger$^{16}$,
B.~Heffron$^{11,13}$,
D.~E.~Jaffe$^{2}$,
S.~Jayakumar$^{3}$,
J.~Koblanski$^{5}$,
P.~Kunkle$^{1}$, 
C.~E.~Lane$^{3}$,
B.~R.~Littlejohn$^{6}$,
A.~Lozano Sanchez$^{3}$
X.~Lu$^{11,13}$, 
F.~Machado$^{6}$,
J.~Maricic$^{5}$,
M.~P.~Mendenhall$^{7}$,
A.~M.~Meyer$^{5}$, 
R.~Milincic$^{5}$, 
P.~E.~Mueller$^{11}$,
H.~P.~Mumm$^{9}$,
R.~Neilson$^{3}$,
X.~Qian$^{2}$,
C.~Roca$^{7}$,
R.~Rosero$^{2}$,
P.~T.~Surukuchi$^{16}$,
F.~Sutanto$^{7}$,
D.~Venegas-Vargas$^{11,13}$, 
P.~B.~Weatherly$^{3}$, 
J.~Wilhelmi$^{16}$,
M.~Yeh$^{2}$,
C.~Zhang$^{2}$ and
X.~Zhang$^{7}$ \\
}


%% file: Introduction.tex

Antineutrinos are generated within an operating nuclear reactor core during $\beta^{-}$ decays of neutron-rich fission fragments.  
Reactors are unique with respect to other terrestrial neutrino sources in generating intense  fluxes with pure electron flavor (\nuebar) and with MeV-scale energies.  
These attributes have enabled reactor-based experiments to perform world-leading three-neutrino-flavor oscillation measurements~\cite{KamLAND_shape, KamLAND_2008,bib:prl_rate,bib:reno,bib:dc,DayaBay:2022orm}.  
Likewise, reactors are uniquely positioned in the global effort to understand the origins of observations that suggest the existence of short-baseline neutrino flavor transformation~\cite{AnomalyWhite,Acero:2022wqg,CHANDLER:2022gvg}.

A simple model often used to fit these anomalies posits an additional sterile neutrino state of roughly eV-scale mass difference with respect to the known neutrino flavors~\cite{giunti_review}, referred to as a `3+1' model.  
This new physics beyond the Standard Model (BSM) is capable of generating deficits in electron-flavor neutrino detection rates with respect to theoretical models at reactor~\cite{bib:mention2011} and radioactive source neutrino experiments~\cite{gallium}, 
\begin{equation}
\label{eq:osc}
P_{\rm{dis}} = \sin^22\theta_{14} \sin^2 \left(1.267 \Delta m^2_{41}({\rm eV}^2) \frac{L({\rm m})}{E_\nu ({\rm MeV})}\right),
\end{equation}
where $\Delta$m$^2_{41}$ is the difference of squared masses, $\theta_{14}$ is a mixing angle describing the electron flavor content of the new mass state, and $E_{\nu}$ and $L$ are the neutrino energy and travel distance (baseline), respectively.  
The new mass state in a 3+1 model could also generate transitions of accelerator-produced muon flavor neutrinos to electron flavor~\cite{LSND:2001fbw,MiniBooNE:2008yuf,MiniBooNE:2020pnu}.  
However, this simple BSM scenario cannot be reconciled with an array of accelerator~\cite{MINOS:2017cae,MicroBooNE:2021rmx,MicroBooNE:2022sdp}, reactor~\cite{DayaBay:2016qvc,bib:neos,danss_osc,STEREO:2022nzk,prospect_prd}, atmospheric~\cite{IceCube_sterile}, and weak decay~\cite{KATRIN:2022ith} experimental results, which has prompted a wide range of subsequent phenomenological investigation~\cite{Acero:2022wqg}.  
%

In the reactor experiment sector, Daya Bay \nuebar flux measurements~\cite{bib:prl_evol} and subsequent developments in nuclear theory~\cite{hayes_evol,bib:fallot2,Perisse:2023efm} and experiment~\cite{Kopeikin:2021rnb} have favored incorrect reactor \nuebar model predictions as the source of observed \nuebar deficits (the `Reactor Antineutrino Anomaly'~\cite{bib:mueller2011}, or RAA) and have reduced community interest in the RAA as a direct indication of BSM physics.  
On the other hand, the Neutrino-4 experiment has recently reported the observation of a deficit with an amplitude similar to the RAA and an $L/E_{\nu}$ character matching that expected from sterile-induced oscillations~\cite{neutrino4_prd}.
In addition, electron flavor deficits continue to appear in MeV-scale radioactive source measurements (termed the `Gallium Anomaly')~\cite{Barinov:2022wfh} and cannot be resolved by the reactor model explanation favored for the RAA~\cite{Giunti:2021kab}.  
Additional high precision data from short-baseline reactor experiments can provide insight on this complex landscape.  

In this Letter, we present new sterile neutrino oscillation results from the PROSPECT-I  reactor neutrino detector deployed at the High Flux Isotope Reactor (HFIR) at Oak Ridge National Laboratory (ORNL).  
Using an improved multi-period selection of inverse beta decay (IBD) $\bar\nu_e + p \to \beta^+ + n$ interaction candidates with lower backgrounds and higher statistical power~\cite{pros_specfinal}, we place limits on sterile neutrino oscillations in previously unaddressed regions between 0.2 and 20~eV$^2$.  

%% file: ExperimentalDescription.tex
The PROSPECT-I detector was located in the HFIR Building at ORNL, adjacent to the reactor pool containment wall in a ground level hallway one story above the reactor.  
The annular cylindrical 85~MW$_\mathrm{th}$ highly enriched uranium HFIR core has a 0.435~m outer diameter and 0.508~m height, compact dimensions that are amenable to measurement of meter-scale oscillation wavelengths associated with 1 -- 10~eV$^2$ active-sterile mass splittings~\cite{VSBL}.  
The PROSPECT-I detector comprised a single stationary \nuebar target spanning distances of 6.7 to 9.2~m from the reactor center.  
The target, filled with $^6$Li-doped organic liquid scintillator (LiLS) exhibiting good pulse shape discrimination (PSD) capabilities~\cite{prospect_ls}, was optically segmented into a rectangular grid of 11\,$\times$14 segments each with 0.145\,$\times$0.145\,$\times$1.18~m$^3$ dimension~\cite{prospect_grid}.  
Long axes of segments were oriented roughly perpendicular to the reactor-detector separation vector to ensure that each segment had a baseline range comparable to the dimensions of the core.  
The detailed experiment geometry and detector design are described in~\cite{prospect_prd,prospect_nim}.  

Scintillation light from charged particle interactions in a segment was detected by one photomultiplier tube (PMT) on each of its two ends.  
Each PMT was encased in a sealed, mineral-oil-filled acrylic housing.   
Following a detector trigger, zero-suppressed 250~MHz digitized PMT waveforms occurring within a common 20~ns arrival time window were recorded and grouped into multi-segment clusters.  
Within a cluster, reconstructed single segment pulses were defined from paired waveforms of PMTs on opposite ends of a segment (double-ended, or DE), or from a single PMT’s waveform (single-ended, or SE) if the segment’s complementary PMT was not operational.
Reconstructed physics quantities, such as energy, position along the segment long axis ($z$), PSD parameter, and segment number, were assigned to each DE pulse.  
The DE reconstruction procedure, described in detail in~\cite{prospect_prd}, produces segment-level energy and position variables that are time-stable over the full PROSPECT data set.  
SE pulse reconstructed quantities were assigned assuming a true $z$-location at the segment midpoint, since timing and amplitude comparisons between segment PMTs could not be performed.  
As a result, SE pulses exhibit poor energy resolution but may offer valuable PSD and segment number information.  
A cluster's reconstructed energy, $E_{rec}$, is defined as the sum of reconstructed energies of all of its DE pulses, while its reconstructed segment number $S_{rec}$ is defined as that hosting the highest-$E_{rec}$ DE pulse.  

This analysis uses PROSPECT-I detector data acquired between March and October 2018, a period encompassing five HFIR fuel cycles and a total of 95.6 reactor-on and 73.1 reactor-off calendar days of data taking.  

\begin{figure}[htbp!]
  \centering
  \includegraphics[trim = 14cm 0cm 9.5cm 0cm, clip=true,  width=0.49\textwidth]{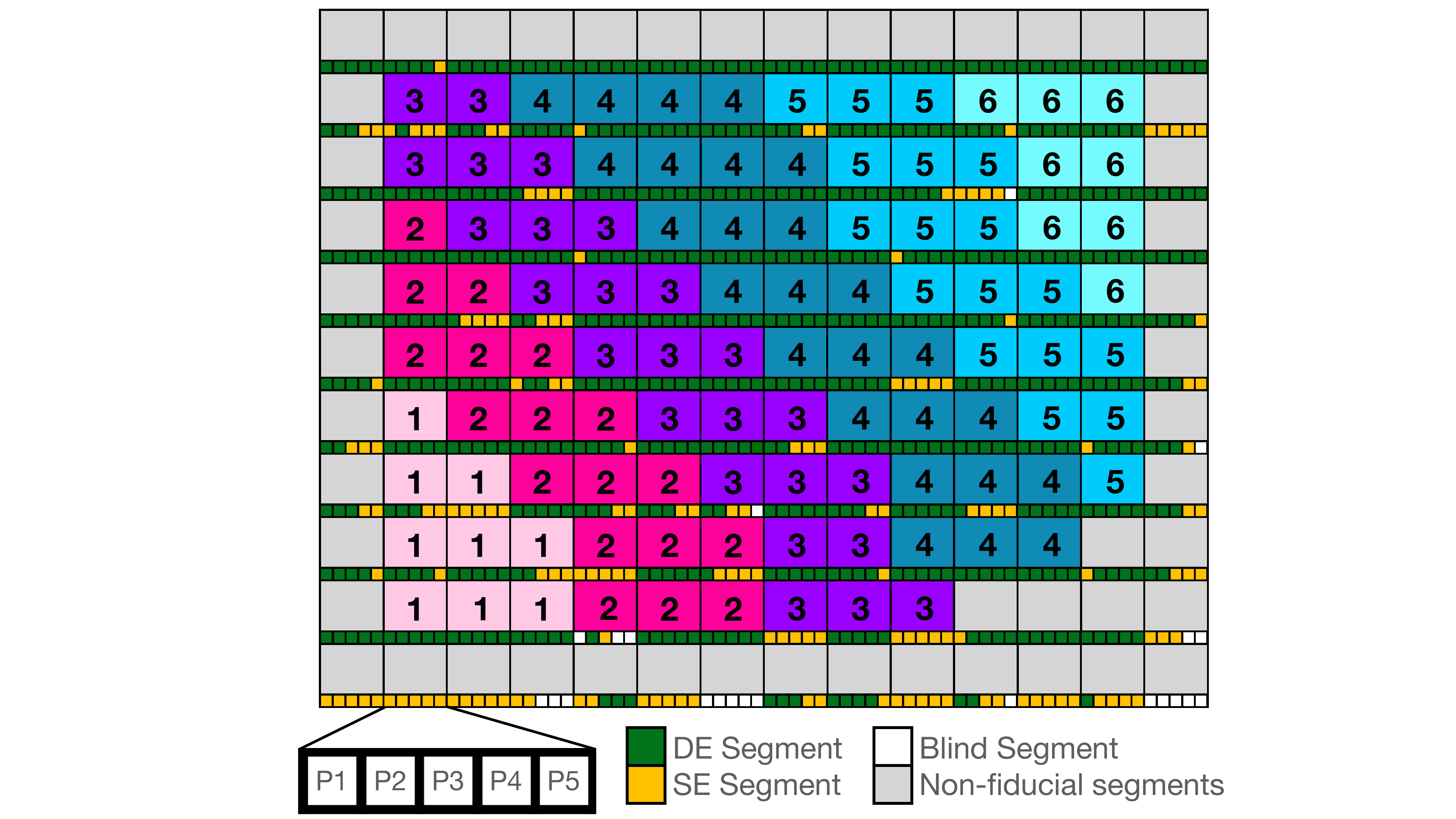}
  \caption{Operational status and baseline binning designation for each PROSPECT detector segment.  The numbers and color of each segment defines its assigned baseline bin, while the colors of each segment's small boxes indicates its operational status for each of the five data taking periods (P1 through P5).  Unnumbered non-fiducial segments are indicated in light gray.}
  \label{fig:baselines}
\end{figure}

As data taking progressed, an increasing number of PMTs experienced current instabilities related to interaction of bare PMT base electronics with LiLS that had leaked into PMT housing interiors; a PMT usually became permanently inoperable within hours or days after first experiencing instability~\cite{Andriamirado:2021qjc}.  
Even with only one functioning PMT, a segment was still capable of generating SE pulses.  
Previous PROSPECT oscillation results used data from only segments that had two fully functional PMTs for the entire data period~\cite{prospect_prd}.  
The analysis reported here uses improved methods, first reported in Ref.~\cite{pros_specfinal}, to address the gradual increase in non-functional PMTs.  
First, the physics data were split into five periods -- each containing one reactor cycle -- that differ only in fully functional and partially functional segment counts, as illustrated in Figure~\ref{fig:baselines}.  
Partially functional (fully non-functional) segment counts in the 154 segment detector varied between periods from as low as 20 (2) to as high as 44 (9).  
Dataset splitting increases PROSPECT's effective exposure and IBD candidate count by enabling consideration of interactions occurring in fully functional segments that became partially functional later in the data taking run.  
In addition, SE pulses from partially functional segments were incorporated into the IBD selection process.  
The re-optimized IBD selection improved the signal-to-background for both cosmic (3.9 vs 1.4) and accidental (4.3 vs 1.8) backgrounds compared to the previous PROSPECT oscillation search~\cite{prospect_prd}.

%% file: AnalysisMethods.tex
Neutrino interaction candidates were selected by exploiting the unique positron ($\beta^{+}$) and neutron ($n$) final state of MeV-scale IBD interactions on protons in the LiLS~\cite{prospect_prd}.  
First, clusters matching the appearance of $n$-$^6$Li captures, which produce energetic highly ionizing pairs of $^4$He and $^3$H nuclei, were identified using tight cuts on segment topology (cluster must contain one DE and zero SE pulses), energy ($E_{rec}$ within 2$\sigma$ of the 0.53~MeV mean reconstructed capture energy), and PSD ($>$2.2$\sigma$ above the electromagnetic band mean).  
Next, a cluster with DE pulses matching the energy (0.8~MeV$<E_{rec}<$7.4~MeV) and PSD ($<$2$\sigma$ above the electromagnetic band mean) characteristics of an IBD $\beta^{+}$ was required in the spatial and temporal vicinity of the $n$-$^6$Li capture: the $\beta^{+}$ cluster should occur in the (-120,-1) $\mu$s range prior to the $n$-capture cluster, with $S_{rec}$ in the same segment (within 140~mm of the capture's $z_{rec}$) or in one of the four adjacent segments (within 60~mm in $z_{rec}$).  
To reduce cosmic fast neutron contamination, $\beta^{+}$ candidate clusters were further required to contain only SE pulses with reconstructed energy $<$0.8~MeV and PSD values $<$3.5$\sigma$ from the electromagnetic band mean.  
IBD candidate $n$-$^6$Li occurring in the temporal vicinity of cosmic muons (0$<t_{IBD}-t_{\mu}<$200~${\mu}$s), neutron-proton recoils  (0$<t_{IBD}-t_{n-p}<$200~${\mu}$s), and other $n$-$^6$Li captures $(-300<t_{IBD}-t_{n-^6Li}<$300~${\mu}$s) were also rejected, along with any IBD-related cluster closely preceded by any cluster of any kind (0$<t_{IBD}-t_{other}<$0.8~${\mu}$s).  

Non-IBD backgrounds from time-correlated cosmogenic signals are statistically subtracted from reactor-on IBD candidate samples using livetime-scaled reactor-off datasets, while backgrounds from non-correlated, accidentally coincident sources are subtracted using an off-window method~\cite{prospect_prd}.  
Subtraction was performed separately for each period.  
This procedure yields a total background-subtracted IBD count of 61029 $\pm$ 338, with counts per data period ranging from 6357 to 16546.  

To perform an oscillation analysis, each data period's IBD candidate set is subdivided into six groups of segments of common baseline ($L$) range, with $L$ defined as the distance from the IBD $\beta^{+}$ cluster's $S_{rec}$ midpoint to the reactor center.  
As shown in Figure~\ref{fig:baselines}, each segment's baseline assignment is consistent across periods.  
IBD candidates are also grouped according to their $\beta^+$ cluster's $E_{rec}$ in 0.2~MeV width bins from 0.8 -- 7.4~MeV.  
This scheme results in a total of 990 ($L$,$E_{rec}$) bins in which the oscillation analysis is performed.  
The presence and analysis treatment of non-functioning segments produces $L$- and period-dependent variations in IBD detection efficiency and $E_{rec}$ response that are modeled with calibration-validated Monte Carlo simulations from the collaboration's \texttt{Geant4}-based PG4 software~\cite{prospect_prd}.  
These response variations are described in more detail in Ref.~\cite{prospect_prd} and in supplementary materials accompanying this manuscript.  

\begin{figure}[htbp!]
  \centering
  \includegraphics[trim = 0cm 0.2cm 1cm 1.2cm, clip=true, width=0.5\textwidth]{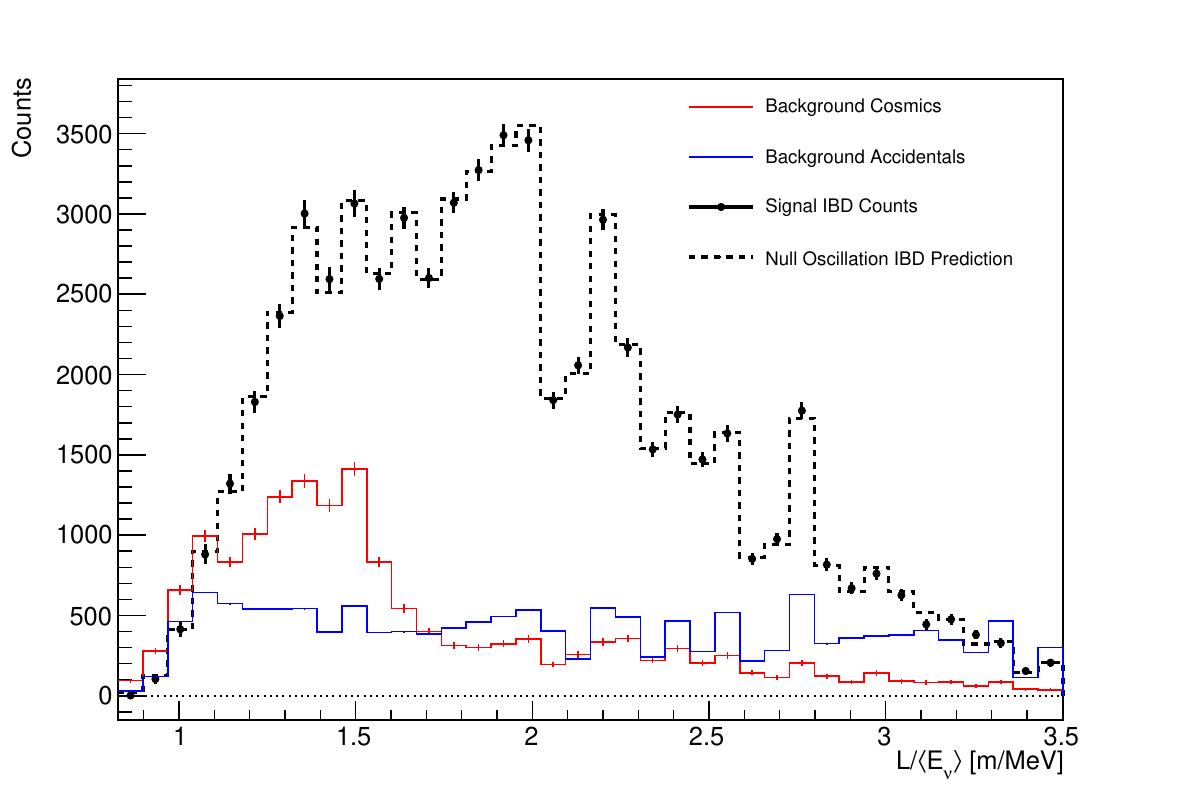}
  \caption{$L$/$\langle E_{\nu}\rangle$ spectrum of background-subtracted IBD signals (black points) and estimated correlated (solid red) and accidental (solid blue) backgrounds during PROSPECT-I reactor-on periods.  $L$ and $\langle E_{\nu}\rangle$ are defined in the text.  The null oscillation IBD signal prediction (dashed black) is also pictured.  Error bars represent 1$\sigma$ statistical uncertainties.}
  \label{fig:lovere1}
\end{figure}

Neutrino oscillations can be visualized by grouping IBD data into bins of common $L/E_{\nu}$.
Following the example of Ref.~\cite{DayaBay:2016ggj}, an average neutrino energy, $\langle$E$_{\nu}\rangle$, corresponding to each PROSPECT-I $E_{rec}$ bin in each period was determined using that period's predicted energy response.  
Figure~\ref{fig:lovere1} shows the background and measured IBD signal (the latter denoted as M$_{l,e}$) for all 990 ($L$,$E_{rec}$) bins regrouped into bins of common $L$/$\langle E_{\nu}\rangle$.  
The center of each ($L$,$E_{rec}$) bin is used to distribute M$_{l,e}$ content in $L$/$\langle E_{\nu}\rangle$ space.  
Structure present in Figure~\ref{fig:lovere1} for both signal and backgrounds arise from the interplay between $E_{\nu}$ spectrum shape, ($L$,$E_{rec}$) binning, and the PROSPECT-I detector geometry.  
The PROSPECT-I dataset begins to have meaningful statistical sensitivity above 1.0\,m/MeV, with average signal-to-background ratios improving from $\sim$2 below 1.5~m/MeV to $>$6 from 1.5 -- 2.75 m/MeV.  
Correlated cosmogenic events represent the largest background to M$_{l,e}$ at the lowest $L$/$\langle E_{\nu}\rangle$ (highest $E_{rec}$), while accidental backgrounds dominate at higher $L$/$\langle E_{\nu}\rangle$ (lower $E_{rec}$).  

Predictions of unoscillated and oscillated PROSPECT-I $E_{rec}$ spectra for each baseline bin (referred to as P$_{l,e}$) and summed over the full detector (P$_{e}$) were generated for each period using a mixture of data-driven and model-based approaches.  
First, the Daya Bay $^{235}$U $E_{\nu}$ model~\cite{Adey:2021rty}, analytically generated true $L$ distributions described in~\cite{prospect_prd}, and $P_{dis}$ as described by Equation~\ref{eq:osc} were used to generate predicted oscillated and unoscillated E$_{\nu}$ spectra for each segment.  
Each period's live time, fully functional segment list, and per-segment detector response matrices were then used to translate truth level distributions into P$_{l,e}$ and P$_{e}$.  
All P$_{l,e}$ and P$_{e}$ bins were then multiplied by a common factor such that the predicted IBD count over all periods matched the measured count, 61029.  
To eliminate dependence of oscillation results on the assumed \uFive~\nuebar~spectrum model, P$_{l,e}$ are multiplied by ($\frac{\mathrm{M}_{e}}{\mathrm{P}_{e}}$), the ratio of baseline-integrated measured and baseline-integrated predicted spectra, prior to being compared to M$_{l,e}$.  
The $L$/$\langle E_{\nu}\rangle$ distribution composed from the null oscillation prediction's P$_{l,e}$ is also pictured in Figure~\ref{fig:lovere1} after applying this ($\frac{\mathrm{M}_{e}}{\mathrm{P}_{e}}$) factor.  Systematic uncertainties in P$_{l,e}$ and P$_{e}$, described later in this manuscript, are much smaller than the statistical uncertainties shown in Figure~\ref{fig:lovere1}.  


\begin{figure}[htbp!]
  \centering
   \includegraphics[trim = 0.0cm 0.2cm 0.5cm 1.0cm, clip=true, width=0.5\textwidth]{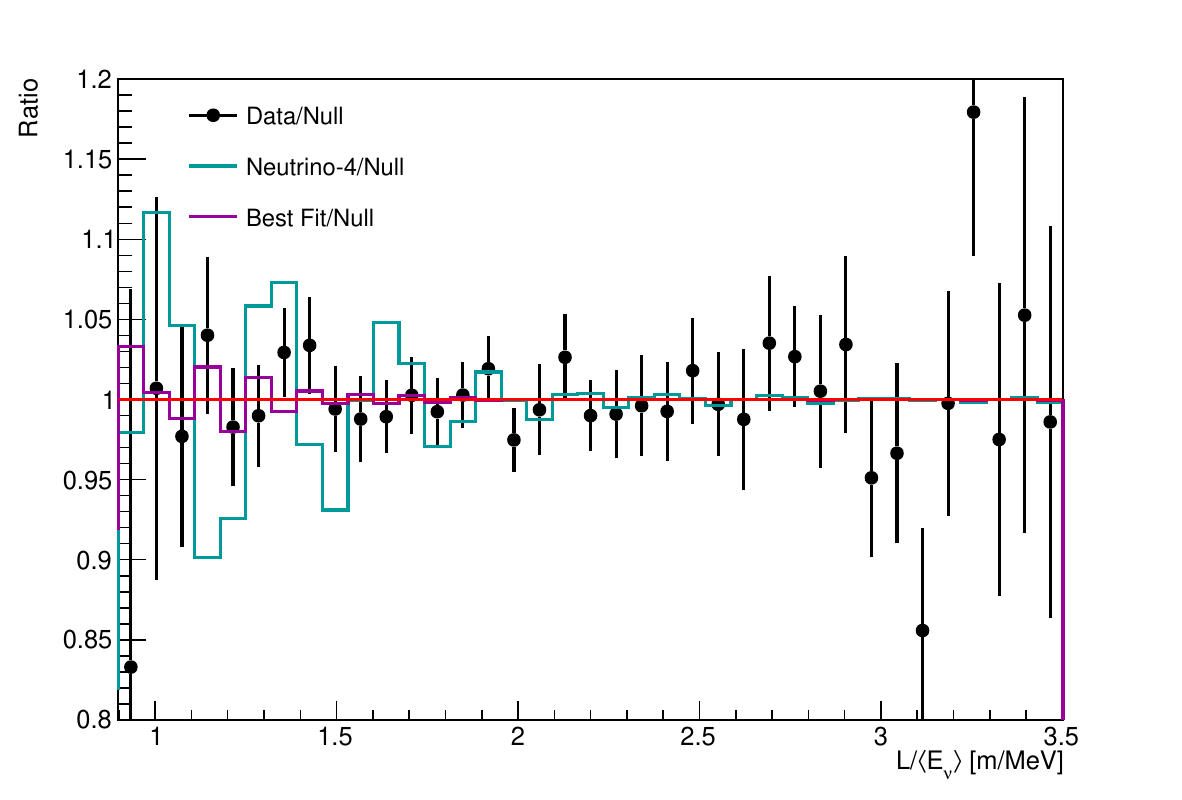}
  \caption{Ratio of $L$/$\langle E_{\nu}\rangle$ features between PROSPECT's IBD signal and its null oscillation prediction.  As described in the text, predicted $\langle E_{\nu}\rangle$ spectra are weighted by the ratio between detector-integrated measured and predicted spectra, ($\frac{\mathrm{M}_{e}}{\mathrm{P}_{e}}$)P$_{l,e}$.  Error bars indicate the statistical uncertainty in the measured IBD spectrum, M$_{l,e}$.  Ratios expected due to oscillations at the PROSPECT data (magenta line) and Neutrino-4 (blue line) best-fit points are also depicted.}
  \label{fig:lovere2}
\end{figure}

Figure~\ref{fig:lovere2} depicts the ratio between measured $L$/$\langle$E$_{\nu}\rangle$ behavior and that predicted by the null oscillation hypothesis.  
As in Figure~\ref{fig:lovere1}, the null oscillation prediction is composed of per-baseline $E_{rec}$ spectra weighted by the ratio of measured to predicted detector integrated spectra, ($\frac{\mathrm{M}_{e}}{\mathrm{P}_{e}}$)P$_{l,e}$.  
Any deviation of the ratio from unity thus reflects $L$-dependent variations in the energy spectrum observed within PROSPECT, as opposed to $L$-independent deviations from a reference $E_{\nu}$ model.  
The predicted ratio is pictured in Figure~\ref{fig:lovere2} for fluctuation-free Asimov M$_{l,e}$ datasets from the best-fit point reported by Neutrino-4, (sin$^22\theta_{14}$,$\Delta$m$^2_{41}$) = (0.36,7.3\,eV$^2$)~\cite{neutrino4_prd}.  
It should be noted that while a non-zero $\theta_{14}$ causes $L$/$\langle$E$_{\nu}\rangle$ oscillations in M$_{l,e}$, it also affects the applied $\frac{\mathrm{M}_{e}}{\mathrm{P}_{e}}$ factor;  thus, a $L$/$\langle$E$_{\nu}\rangle$ representation with multiple ($L$,$E_{rec}$) bins of differing $\frac{\mathrm{M}_{e}}{\mathrm{P}_{e}}$  per $L$/$\langle$E$_{\nu}\rangle$ bin will exhibit suppressed oscillation features relative to the original ($L$,$E_{rec}$) representation.  
This suppression is minimal for oscillation wavelengths smaller than the detector size, as is the case for Neutrino-4’s most favored oscillation parameters, for which oscillatory features are resolvable below 2.5~m/MeV. Damping at higher L/$\langle$E$_{\nu}\rangle$ is primarily caused by energy smearing from IBD positron and annihilation gamma energy leakage into non-active detector regions and secondarily by baseline smearing from the reactor's finite size.  
PROSPECT-I’s IBD data is qualitatively consistent with the null hypothesis of no oscillations.  

%% file: Results.tex
A Combined Neyman Pearson (CNP)~\cite{pros_stats} test statistic was used to quantify the level of agreement between data and oscillated and unoscillated predictions in ($L$,$E_{rec}$) space: 
\begin{equation}
\label{eq:oscchi2}
\chi^2_{CNP} = \bm{\Delta^{\textrm{T}}}\textrm{V}_{\textrm{tot}}^{-1}\bm{\Delta}.
\end{equation}
The 990-element vector $\bm{\Delta}$ represents the data-prediction agreement in per-baseline $E_{rec}$ spectra~\cite{prospect_prd}, with each period contributing 198 elements, 
\begin{equation}\label{eq:delta}
\Delta_{l,e} = M_{l,e}- \frac{M_{e}}{P_{e}}P_{l,e}. 
\end{equation}
As in Figure~\ref{fig:lovere2}, since the predicted $E_{rec}$ spectrum at each baseline $l$ in $P_{l,e}$ is only compared to $M_{l,e}$ after correcting for the relative difference in $P_{e}$ and $M_{e}$ for each period, $\Delta_{l,e}$ is not affected by well-known biases in reactor E$_{\nu}$ models~\cite{bib:reno_shape,bib:prl_reactor,bib:neos,bib:prosDBjoint,bib:prosSTEREOjoint,pros_specfinal}.  
The covariance matrix V$_{\textrm{tot}}$ describes statistical and systematic uncertainties and their correlations between analysis bins.  
Statistical uncertainties are calculated as defined by the CNP test statistic, which suppresses biases related to low per-bin statistics.  
Systematic uncertainties related to absolute and/or baseline uncorrelated energy scales, energy resolution, energy scale non-linearity, energy leakage into non-active detector volumes, background and IBD signal normalizations, and reactor-detector baselines are calculated using Monte Carlo or analytic methods described in detail in~\cite{prospect_prd}.  
As in PROSPECT's other multi-period  analysis~\cite{pros_specfinal}, signal and detector response (background) systematics are treated as correlated (uncorrelated) between periods.  
Oscillation sensitivity is dominated by statistical uncertainties over most of the tested parameter space.  
The primary systematic impacting the analysis is related to volume and IBD detection efficiency variations between segments; each segment is assigned an uncorrelated IBD rate normalization uncertainty of 5\%, with this value primarily based on the level of segment-to-segment consistency in detection rates of uniformly distributed $^{219}$Rn-$^{215}$Po decays in the LiLS.  
This systematic, which has a much larger sensitivity impact than all other considered  sources, plays a modest role at $\Delta$m$^2<$1\,eV$^2$.  

In a scan over (sin$^22\theta_{14}$,$\Delta$m$^2_{41}$) oscillation phase space, a minimum $\chi^2_{CNP,min}$ of 827 is identified for the PROSPECT-I dataset, at (0.421,15.2 eV$^2$).  
Predicted $L$/$\langle E_{\nu}\rangle$ behavior at this point is illustrated in Figure~\ref{fig:lovere2}.
The $\chi^2_{CNP,min}$ value is 2.57 units lower than that obtained for the null oscillation case (sin$^22\theta_{14}$=0).  
A frequentist compatibility test was then performed by generating 2000 toy PROSPECT-I unoscillated datasets with expected statistical and systematic fluctuations applied via Cholesky decomposition of V$_{\textrm{tot}}$.  
The $\Delta \chi^2_{CNP}$ of 2.57 between minimum and null points was found to be larger than 13\% of unoscillated toys, indicating consistency of PROSPECT-I's data with an absence of sterile neutrino oscillations.  
A similar test was applied to the Neutrino-4 best fit point (sin$^22\theta_{14}$,$\Delta$m$^2_{41}$)=(0.36,7.3 eV$^2$).  
Its $\Delta \chi^2_{CNP}$ of 35.16 with respect to the PROSPECT best-fit point is roughly two times larger than any of the 2000 toys thrown at this grid point, indicating its strong ($>$3$\sigma$) incompatibility with PROSPECT's data.

\begin{figure}[tbhp!]
  \centering
    \includegraphics[trim = 0.5cm 0cm 0cm 0cm, clip=true, width=0.5\textwidth]{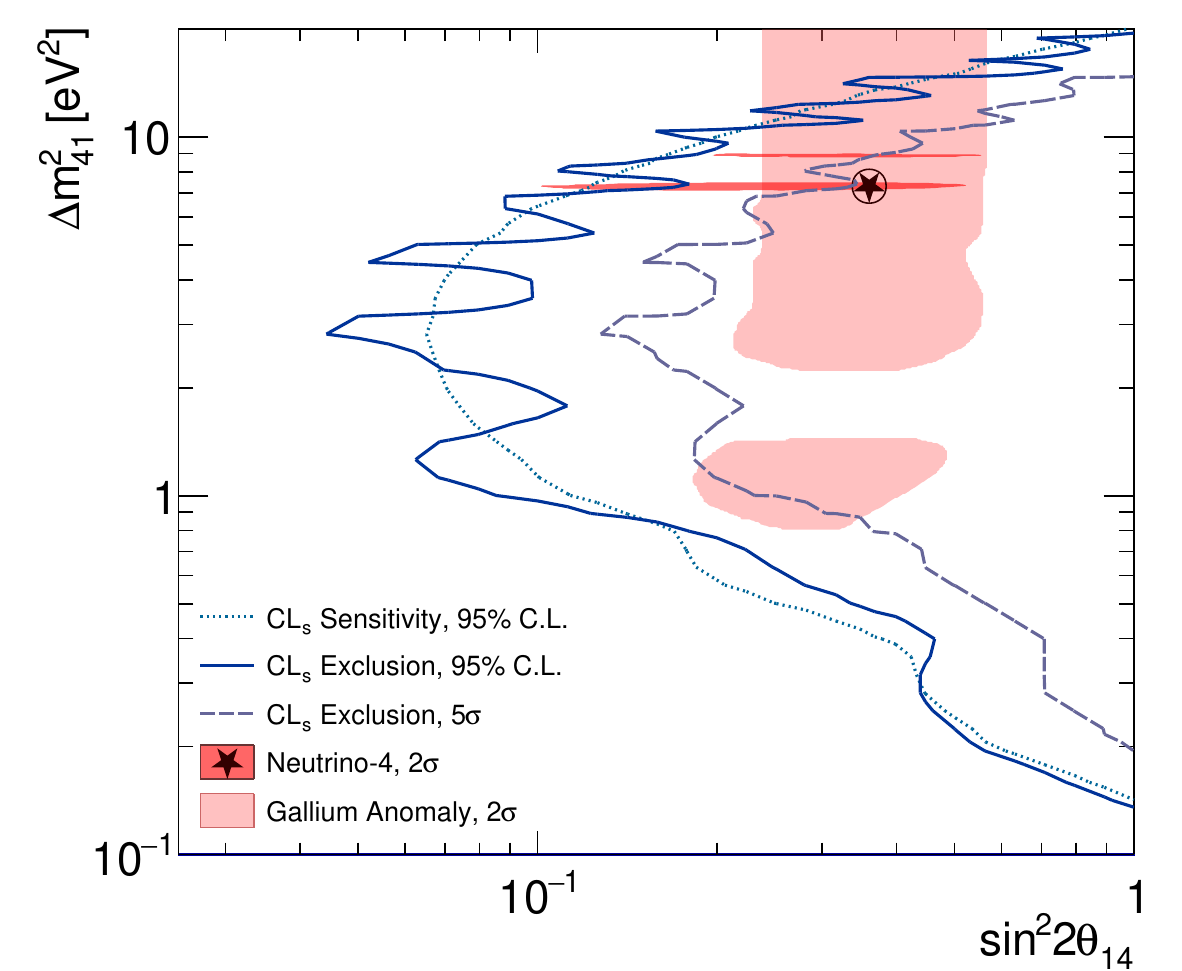}
    \caption{Phase space for 3+1 sterile neutrino oscillations excluded by the multi-period PROSPECT-I dataset.  Exclusion (solid) and sensitivity (dotted) curves, presented for 95\% confidence level (CL), were generated using the Gaussian CLs method.  Exclusion (dashed) at 5$\sigma$ CL is also shown. Shaded suggested phase space regions from the Gallium Anomaly~\cite{Barinov:2022wfh} (light pink) and Neutrino-4~\cite{neutrino4_prd} (red) are also pictured.}
  \label{fig:phasespace}
\end{figure}

An exclusion contour was generated using the Gaussian CL$_{s}$ method~\cite{cls}.  
Obtained 95\% confidence level (CL) and 5$\sigma$ exclusion contours are pictured in Figure~\ref{fig:phasespace} along with the expected median sensitivity for this dataset, which is determined by substituting the null oscillation Asimov dataset in place of PROSPECT-I measurements.  
Most of Neutrino-4's suggested region appears to the right of the 95\% CL PROSPECT-I exclusion contour, and the Neutrino-4 best-fit point is excluded at more than 5$\sigma$ CL.  
The 95\% CL PROSPECT-I contour excludes regions of 3+1 sterile neutrino parameter space above 3~eV$^2$ previously unexplored by other neutrino experiments~\cite{bib:neos,danss_osc,STEREO:2022nzk,MicroBooNE:2021rmx,KATRIN:2022ith}, including all space below 10~eV$^2$ suggested at 2$\sigma$ CL by BEST and other Gallium Anomaly experiments~\cite{Barinov:2022wfh}.  

In summary, we have performed a new 3+1 sterile oscillation analysis with  multi-period IBD data from the PROSPECT-I detector deployed at the compact HFIR research reactor.  
Thanks to an improved methodology, these data contain lower backgrounds and higher IBD statistics than PROSPECT’s previous analysis~\cite{prospect_prd}, enabling greater sensitivity to neutrino oscillation.  
Visualizing PROSPECT data in $L/E_{\nu}$ space, we see no obvious appearance of high amplitude short-baseline neutrino oscillation  previously reported by the Neutrino-4 experiment.  
When binned in ($L$,$E_{rec}$) space, the data is found to be typical of that expected in the absence of short-baseline oscillation and compatible with null oscillation (p-value 0.87).  
PROSPECT's data excludes new high-$\Delta$m$^2_{14}$ regions of phase space at greater than 95\% CL, including most of the Neutrino-4 suggested space and previously unexplored regions suggested by the recently strengthened Gallium Anomaly.   